%% file: main.tex
\documentclass[runningheads]{llncs}

\usepackage[T1]{fontenc}

\usepackage{hyperref}
\usepackage{xurl}
\usepackage{graphicx}

\usepackage{color}

\hypersetup{ 
  colorlinks   = true, 
  allcolors   = blue, 
}

\usepackage{booktabs}
\usepackage{multirow}
\usepackage{svg}
\usepackage{siunitx}
\begin{document}

\title{Performance Testing of ChaCha20-Poly1305 for Internet of Things and Industrial Control System Devices}

\titlerunning{Performance Testing of ChaCha20-Poly1305 for IoT and ICS Devices}

\author{Kristján~Orri~Ragnarsson\inst{}\orcidID{0009-0000-6377-3113} \and
    Jacky Mallett\inst{}\orcidID{0000-0003-1370-996X}}

\authorrunning{K. O. Ragnarsson \and J. Mallett}

\institute{Reykjavik University, Reykjavik, Iceland\\ \email{\{kristjanr20,jacky\}@ru.is}}

\maketitle

\begin{abstract}
    Industrial Control Systems (ICS), and many simple Internet of Things (IoT) devices, commonly communicate using unencrypted or unauthenticated protocols.
    For ICS this is an historical carryover since the introduction of these systems predated practical lightweight cryptography. As the processing power of small devices has grown exponentially at the same time as new, more efficient encryption algorithms have become available, end device encryption of communication protocols is becoming much more practical, but is still not widely used with ICS protocols such as Modbus and IEC61850 (GOOSE) which have tight requirements for both latency and variance. Newer micro-processors can also present challenges both to measurement and use, since features such as dynamic frequency scaling can significantly impact performance measurements.
    In this paper, we measured the time cost of adding encryption into the communication cycle of low-cost edge devices using ChaCha20-Poly1305, and show that in the worst case the encryption cycle took less than 6 percent of the latency requirements of Goose, and less than 3\% for IEC-60834-1 on Raspberry Pi 4, and an Intel N95 Mini PC, which is well within the specified latency requirements for these protocols.

\end{abstract}

\section{\label{sec:introduction}Introduction}
In recent years, cyber attacks targeting critical infrastructure have highlighted vulnerabilities in ICS and supervisory control and data acquisition (SCADA) networks. Notable incidents, such as the 2015 power outage in Kyiv affecting 225,000 residents \cite{polityukUkrainesPowerOutage2017}, and more recent attacks that disabled tens of thousands of ICS sensors in Russia~\cite{noauthor_moscollector_2024}, have demonstrated the growing risks facing these systems. Although some attacks have been state-sponsored and linked to geopolitical conflicts, critical infrastructure is also increasingly being targeted by criminal groups and advanced persistent threat(APT) actors. For example, the 2021 ransomware attack on the Colonial Pipeline disrupted fuel distribution in the eastern United States~\cite{u_s_government_accountability_office_colonial_2025}, and incidents such as Industroyer and TRITON demonstrate how ICS-specific malware can be used to manipulate or disable safety and control systems~\cite{giles_triton_2019,zafra_industroyerv2_2022}.

A common vulnerability in ICS and SCADA environments is the use of unencrypted or unauthenticated communication protocols~\cite{alanazi_scada_2023}. This largely stems from historical factors. The programmable logic controller (PLC) was invented in the 1960s, and replaced hardwired relays in physically secure locations where encryption was arguably not necessary, but was also not feasible at that time given the computational power available to these systems~\cite{noauthor_evolution_2021}. Despite decades of technological advancements the perception persists that ICS devices are not able to use encryption due to the real-time performance requirements that they have to operate within for the processes they are controlling~\cite{gumaei_robust_2020}.

With the explosive increase in computing power over the last decades and also improvements in the computational efficiency of modern ciphers, interest has grown in Lightweight Cryptographic Algorithms suitable for ICS and IoT devices~\cite{dhanda2020}.   To explore these issues we examine here the specific question of whether the peculiar constraints on ICS and real time IoT devices are still an issue in preventing encryption, or whether the efficiency gains of improved algorithms and improved hardware can keep encryption time within the time limits imposed by ICS protocols such as IEC61850 (GOOSE). By modern standards some of these latency requirements are no longer as heinous as they once were. For example, the Goose protocol requires $<$ \qty{4}{\milli\second} latency, IEC 60834-1 requires $<$ \qty{10}{\milli\second} for inter tripping protection commands and  SCADA operates with $<$ \qty{1}{\second} latency requirements~\cite{chelluri_integration_2011}.

For this purpose, we chose to explore the performance impact of ChaCha20-Poly1305 authenticated encryption on data payloads similar to those used for ICS communications. ChaCha20 is a performant encryption cipher derived from Salsa20~\cite{Bernstein2008,bernstein_chacha_2008},  while Poly1305 is a message authentication code~\cite{bernstein_poly1305-aes_2005}. We evaluated whether this combination offers a viable balance between security and performance in resource-constrained environments on both a Raspberry Pi 4 (Model B, Rev 1.1) and an Intel N95 Mini PC, and observed performance times that were well within the requirements for IEC 60834-1.

In the rest of this paper we discuss related works and background~\ref{sec:related_works}, section~\ref{sec:methods} describes the methods and design used and  section~\ref{sec:results} presents our results. Section~\ref{sec:discussion} provides a discussion, and section~\ref{sec:conclusion_future_works} concludes the paper and outlines future work.

\section{\label{sec:related_works}Related Works and Background}
Security concerns in ICS mirror, but are distinct to those of IT security. The well known Confidentiality, Integrity and Availability (C.I.A.) triad is often regarded as being inverted in this area, with reliable availability being regarded as the single most important objective for critical infrastructure such as power stations~\cite{saad_review_2019}.

Cyber attacks that are addressed by encryption fall into two general categories: attacks on confidentiality and attacks on integrity. Intercepting plain text packets is an example of a confidentiality attack. For example, malware might use intercepted communications for network reconnaissance to identify nodes to attack. In contrast, fabricating data to force an incorrect action is an integrity attack. An example would be malware sending disconnect commands to breakers to cause a blackout. Encryption prevents malware from exploiting unencrypted traffic for reconnaissance or data injection,  provided that the malware is unable to use infected hosts to attack the application performing the encryption, and cannot intercept the keys required to encrypt and decrypt the traffic. What can be said, is that if traffic is not encrypted, any machine able to eavesdrop on the communication channels being used, can potentially perform a man-in-the-middle attack, and modify or inject commands and data.

Fauri et al. argue however that confidentiality is not critical in ICS environments. They emphasize the importance of availability and integrity, and claim that encryption has no place in ICS networks and could even be detrimental, as no known real-world attacks would have been prevented by encryption. For example in the Stuxnet attack, the attack proceeded against the software issuing the commands(i.e. the application layer of the network), rather than against the protocol itself. Instead, encryption may hinder other crucial defenses, such as intrusion detection systems (IDS) and network troubleshooting. According to their view, the benefits of encryption can be achieved through message authentication alone, without the need for full encryption~\cite{fauri_encryption_2017}.

This argument ignores both the evolving nature of malware, and issues around network layer man-in-the-middle attacks. Just because malware utilizing a certain technique has not been seen does not mean it will not be made in the future, or has already been made and simply not detected. Intrusion detection systems are not particularly hard to circumvent, and next-generation firewalls (NGFWs) can also decrypt network traffic for deep packet inspection which negates the argument that encryption hinders IDS~\cite{wilson_importance_2024}.  Injection attacks against TCP/IP connections are straightforward, even more so with UDP, and so unencrypted protocols can be regarded as considerably increasing the range of the attack surface and to far more computers than just the SCADA host used in Fauri's example.

Encryption in ICS applications must however be highly performant due to the strict latency requirements of these systems. Added delays from encryption must remain within the real time limits defined for the protocol using it,  so as not to compromise operational security.

Alves et al. demonstrated the feasibility of integrating end-to-end encryption into ICS by implementing AES-256 in OpenPLC, an open-source programmable logic controller platform~\cite{alves_securing_2017}. Their work showed improvements in communication confidentiality and integrity, and they analyzed the impact on system throughput. However, they did not report on latency or computational overhead, and the lack of published source code limits reproducibility and further research. Razzaque et al. analysed a number of protocols based on the number of clock cycles required for their instruction set, however this approach assumes that the execution of those instructions is identical on all hardware, which as we will report is not always the case~\cite{razzaque.2021}.

Upadhyay et al. proposed a hardware simulation testbed for SCADA systems and evaluated the performance of several encryption algorithms in terms of execution time and memory usage. Their testbed consisted of a Raspberry Pi as a remote site, a Windows machine running VTSCADA as the control center, and an MQTT server as the WAN connection. They showed that encrypting messages through the MQTT server with various techniques: Prime Counter \& Hash Chaining (PCHC), Ascon algorithm using compression rate (ACR), and American Gas Association Report 12 (AGA-12)—resulted in latencies as low as \qty{0.72}{\milli\second} for PCHC, \qty{1.34}{\milli\second} for ACR, and \qty{4.64}{\milli\second} for AGA-12~\cite{UPADHYAY2024100705}.

This report builds upon such works, particularly the performance analysis by Upadhyay et al., but aligns more with the software-centric approach of Alves et al., where encryption is implemented at the software level using either message encryption or TLS-secured connections and then its performance is measured whilst running. For this, we examine the performance of two recent encryption algorithms, ChaCha20 and Poly1305.

ChaCha20 is a variant of Salsa20 that improves diffusion through reordered operations and altered rotation constants, while maintaining the same number of rounds and operands. It also reduces register use, which improves performance on many CPUs while preserving its parallelizability~\cite{bernstein_chacha_2008}. Salsa20 is a family of 256-bit stream ciphers designed for efficiency and parallelization across various processor architectures. It relies solely on addition, exclusive-or, and rotation operations that are uniformly fast across most CPUs, unlike integer multiplication which varies depending on implementation. Salsa20 generates 64-byte blocks using a 32-byte key, an 8-byte nonce, and an 8-byte block counter. Each block is independent, making the stream both randomly accessible and parallelizable~\cite{Bernstein2008}.

Poly1305 is a message authentication code that generates a 16-byte tag using a 32-byte key and a 16-byte nonce~\cite{bernstein_poly1305-aes_2005,nir_chacha20_2015}. Originally designed for use with AES, it has since been adopted in conjunction with other ciphers, most notably ChaCha20. The combination ChaCha20-Poly1305 is widely used in protocols such as IPsec, TLS, and OpenSSH~\cite{nir_chacha20_2015_ipsec,langley_chacha20-poly1305_2016,miller_chacha20_nodate}.

\begin{figure}[htb]
    \centering
    \includesvg[width=1\linewidth]{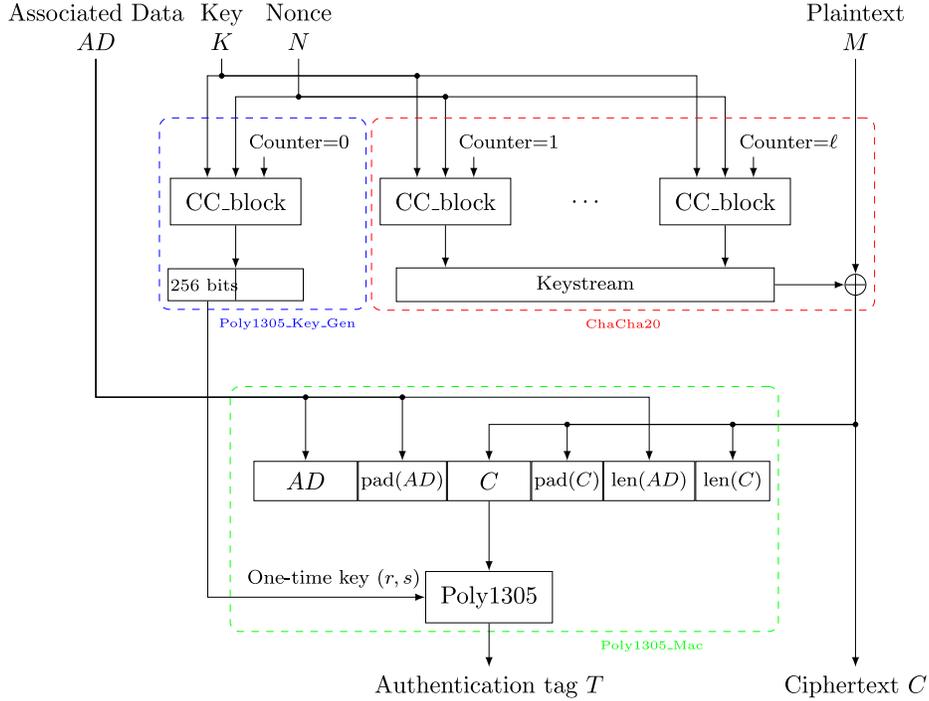}
    \caption{Diagram of ChaCha20-Poly1305 by Morz25, CC BY-SA 4.0, via Wikimedia Commons~\cite{morz25_ChaChaPoly_diagram}}
    \label{fig:svg}
\end{figure}

AES-NI is an instruction set introduced by Intel in 2010 to improve the performance of AES by adding full hardware support for AES operations, thus reducing complex operations to simple instructions for the procedures~\cite{gueron2010intel}.  Chesebrough and Conlon found CyaSSL\footnote{Old name for WolfSSL} achieved \SI{579}{\mega\byte\per\second} when using AES-NI compared to \SI{170}{\mega\byte\per\second} without it~\cite{chesebroughImplementationPerformanceAESNI2012}.

ChaCha20-Poly1305 is not unexplored in ICS, Jingran et al. implemented encrypted Modbus/TCP transmission using ChaCha20-Poly1305 TLS, however they did not report on their experimental setup beyond that it was done on the PC platform, nor did they analyse the affect this experiment had on the performance of the system or the code used during the experiment~\cite{jingranResearchImplementationSecure2020}. As we detail below, CPU frequency scaling and other features of modern CPU´s can make performance measurements difficult to calibrate and also raise questions about the reliability of performance of some IoT devices using this hardware.

Furthermore, Niu et al. evaluates direct on device encryption and encryption via a proxy server, using the ChaCha20-Poly1305 stream cipher encrypting the data section of packets, thus not using TLS to avoid handshake overhead~\cite{niuExaminingSuitabilityStream2024}. However, they fail to give concrete numbers in their latency evaluation on anything other than the added latency of on device encryption which they find to be \qtyrange{0.1}{0.2}{\milli\second}.

\section{\label{sec:methods}Methods}

\subsection{System Setup}
The experiments were run on a Raspberry Pi 4 Model B Rev 1.1 with \qty{4}{\giga\byte} of RAM, running Raspberry Pi OS (kernel: \texttt{6.6.31+rpt-rpi-v8 \#1 SMP PREEMPT Debian}) for arm performance and an Intel N95 Mini PC with \qty{12}{\giga\byte} of RAM running Debian Linux with a real-time kernel patch (kernel: \texttt{6.1.0-37-rt-amd64 \#1 SMP PREEMPT\_RT Debian}) for x86\_64 performance.

\subsection{Cipher Implementation}
The ChaCha20-Poly1305 cipher used in this experiment is implemented by the WolfSSL library (version 5.7.6, commit \texttt{1c56a2674a5ef51c}). The library was statically compiled into the encryption/decryption binary to ensure fairness in all measurements.

The WolfSSL library is built with custom \texttt{user\_settings.h} to remove unnecessary ciphers and functionality. ChaCha20-Poly1305 is implemented in our binary using \texttt{wc\_ChaCha20Poly1305\_Encrypt} and \texttt{wc\_ChaCha20Poly1305\_Decrypt} from WolfSSL.

\subsection{Measurements}

Measurements pose an interesting challenge when trying to figure out the effect encryption has on performance. There are many intricacies that can affect the measurement, such as the behaviour of blocking versus non-blocking i/o functions with different behaviour between terminal programs. For example, \texttt{write()} calls on Terminals vs. Virtual Terminals (VT): calling \texttt{write()} on a terminal will block until the message has displayed on screen, while on a VT it is only blocking until it has been copied into the correct buffer. This can non-obviously increase runtime considerably, we saw a \qtyrange{100}{2000}{\percent} increase in runtime on a TTY due to this~\cite{kostixAnswerProgramSIGNIFICANTLY2017}. Dynamic frequency scaling such as Intel Turbo boost allows CPUs to run at a base frequency and then scale  up on the fly when the workload requires it~\cite{WhatIntelTurbo}, this was observed during testing where a script running the tests performed better than running each by hand in the console as the script increased the workload sufficiently to trigger the use of a higher CPU frequency.

With this in mind, our test was performed using a binary that performs the encryption and times it using the \texttt{timespec} function, and then returns the execution time of different components of the encryption process via \texttt{stdout}. A combination of bash scripts and a go program are used to run the binary in varying configurations for different measurements. We stored the data from the binary in a variable to avoid the variable overhead of the program caused by writing to \texttt{/dev/tty} slowing down the program with unnecessarily slow blocking calls. Data from external testing tools \texttt{perf} and \texttt{time} that output to \texttt{stderr} are separated and piped into \texttt{/dev/shm} buffers that are stored in ram.

During testing, it was observed that the \texttt{timespec} counter overflows \numrange{4}{28} times during \num{100000} test runs, resulting in a negative time during those runs, since the time to do something cannot be negative, these data points were removed prior to analysis.

The code used during the making of this paper is publicly accessible via our GitHub repository\footnote{https://github.com/Kristjan-O-Ragnarsson/Cha-Chop}.

\section{\label{sec:results}Results}
Tables~\ref{tab:chacha_time_arm}\textendash\ref{tab:chacha_time_x86_lock} show the times spent in different parts of the ChaCha20-Poly1305 encryption and decryption process, With Table \ref{tab:chacha_time_arm} showing results from the Raspberry Pi, Table~\ref{tab:chacha_time_x86} the Mini PC with no frequency restriction and Table~\ref{tab:chacha_time_x86_lock} the Mini PC restricted to its base frequency of \qty{800}{\mega\hertz}.

\textbf{Functional time} measures the duration from random generation to the end of cryptographic functions. This more accurately reflects the time spent on encryption and decryption in the overall communication cycle.

\textbf{Random time} shows the time spent on random generation for nonce.

\textbf{Encryption time} shows the time spent reading plaintext encrypting and constructing the final message (ciphertext, nonce, tag) and copying it to a single buffer.

\textbf{Decryption time} is reading the message from file, then splitting it into ciphertext, nonce and tag, then decrypting and printing the message.

\input{new_pi}

\input{new_x64}

\input{new_base}

Functional time is \qtyrange{155}{163}{\micro\second} for the Raspberry Pi, the frequency locked Mini PC is at a similar place with \qtyrange{164}{172}{\micro\second}, thus for Goose the Raspberry Pi uses \qtyrange{3.875}{4.075}{\percent} and the Mini PC \qtyrange{4.075}{4.3}{\percent} of the maximum latency requirements and \qtyrange{1.55}{1.63}{\percent} (Raspberry Pi) and \qtyrange{1.63}{1.72}{\percent} (Mini PC) in IEC 60834-1.

Looking at 5th–95th percentile range, the Raspberry Pi spans \qtyrange{135}{224}{\micro\second}, corresponding to \qtyrange{3.375}{5.6}{\percent} of the Goose requirement and \qtyrange{1.35}{2.24}{\percent} of IEC 60834-1. The frequency locked Mini PC ranges \qtyrange{147}{207}{\micro\second} corresponding to \qtyrange{3.675}{5.175}{\percent}  and \qtyrange{1.47}{2.07}{\percent} respectively.

With Dynamic frequency scaling enabled, the Mini PC achieves a functional time of \qtyrange{57}{60}{\micro\second},  with a 5th–95th percentile range of \qtyrange{52}{64}{\micro\second}, therefore \qtyrange{1.425}{1.5}{\percent} of the maximum latency requirements for Goose and \qtyrange{0.57}{0.6}{\percent} of IEC 60834-1. The 5th–95th percentile range corresponds to \qtyrange{0.13}{0.16}{\percent} of GOOSE and \qtyrange{0.52}{0.64}{\percent} of IEC 60834-1.

Table~\ref{tab:chacha_mem_arm} shows that the maximum resident set size remained consistent across input sizes staying consistently at \qty{1152}{\kilo\byte} on the Raspberry Pi. This indicates that the encryption process is not the biggest usage of memory within this range of plaintext sizes. However, on the Mini PC the maximum resident set size fluctuated from \qtyrange{1372}{1544}{\kilo\byte} averaging around \qty{1434}{\kilo\byte}, as seen in Table~\ref{tab:chacha_mem_x86}.

\begin{table}[!ht]
    \centering
    \caption{Maximum resident set size for the ChaCha20-Poly1305 encryption module with plaintext ranging from \qtyrange{28}{224}{\byte} on the Raspberry Pi (AArch64)}
    \begin{tabular}{l c }
        \toprule
        Message Size     & Peak Memory Usage      \\
        \midrule
        \qty{28}{\byte}  & \qty{1152}{\kilo\byte} \\

        \qty{56}{\byte}  & \qty{1152}{\kilo\byte} \\

        \qty{112}{\byte} & \qty{1152}{\kilo\byte} \\

        \qty{224}{\byte} & \qty{1152}{\kilo\byte} \\

        \bottomrule
    \end{tabular}

    \label{tab:chacha_mem_arm}
\end{table}

\begin{table}[!ht]
    \centering
    \caption{Maximum resident set size range and average over 7 runs for the ChaCha20-Poly1305 encryption module with plaintext ranging from \qtyrange{28}{224}{\byte} on the Mini PC (x86-64)}
    \begin{tabular}{l c c}
        \toprule
        Message Size     & Peak Memory Usage                 & Avg. Peak Memory Usage \\
        \midrule
        \qty{28}{\byte}  & \qtyrange{1372}{1480}{\kilo\byte} & \qty{1424}{\kilo\byte} \\

        \qty{56}{\byte}  & \qtyrange{1408}{1544}{\kilo\byte} & \qty{1464}{\kilo\byte} \\

        \qty{112}{\byte} & \qtyrange{1380}{1480}{\kilo\byte} & \qty{1428}{\kilo\byte} \\

        \qty{224}{\byte} & \qtyrange{1372}{1544}{\kilo\byte} & \qty{1419}{\kilo\byte} \\

        \bottomrule
    \end{tabular}

    \label{tab:chacha_mem_x86}
\end{table}

\begin{table}[!ht]
    \centering
    \caption{Binary size of the ChaCha20-Poly1305 encryption module with debug symbols, normal and none across AArch64 and x86-64}
    \begin{tabular}{l c c c}
        \toprule
        ISA     & Debug Symbols        & Normal               & Stripped             \\
        \midrule
        AArch64 & \qty{85}{\kilo\byte} & \qty{71}{\kilo\byte} & \qty{67}{\kilo\byte} \\
        x86-64  & \qty{35}{\kilo\byte} & \qty{28}{\kilo\byte} & \qty{19}{\kilo\byte} \\
        \bottomrule
    \end{tabular}

    \label{tab:binary_size}
\end{table}

\noindent The size of the test binary varied in size depending on symbols and architecture, achieving the smallest of \qty{19}{\kilo\byte} stripped\footnote{Linux command \texttt{strip} removes symbols from executable binaries reducing the size of the file} of symbols on x86-64 as seen in Table~\ref{tab:binary_size}.

The execution time was measured using \texttt{perf}\footnote{\texttt{linux-perf} on Debian \texttt{apt}} and \texttt{timespec} on the following components:
\begin{itemize}
    \item Time spent on random number generation (e.g., nonce generation),
    \item Time to copy the message into the input buffer,
    \item Time taken by the encryption or decryption operation itself,
    \item Time to copy the output to its respective buffer.
\end{itemize}

This timing setup is intended to reflect realistic processing costs in constrained environments, capturing both cryptographic operations and necessary memory operations.

Peak memory usage was measured using \texttt{time -v}\footnote{GNU Time is required for the -v argument}, specifically using the \textit{Maximum resident set size} metric, which represents the maximum physical memory used by the process.

The binary size was measured in bytes. All binaries are statically compiled with their required libraries (e.g., WolfSSL) to ensure a fair comparison across systems where these libraries may not be available by default.

\section{\label{sec:discussion}Discussion}
The memory footprint on the Raspberry Pi remained constant at \qty{1152}{\kilo\byte} while the Mini PC fluctuated around \qty{1434}{\kilo\byte} for all the sizes of plaintext tested (\qtylist{28;56;112;224}{\byte}). This consistency shows predictability, which is good for resource-constrained devices and real-time devices where memory budgeting can be critical.

All devices show an upward trend for execution time with regard to message size, the Raspberry Pi mean increased by \qty{4.1}{\micro\second} while the 95th percentile increased by \qty{10.5}{\micro\second}, while the Mini PC with its frequency locked showed a mean increase of \qty{8.5}{\micro\second} and \qty{22}{\micro\second} (95th percentile). With dynamic frequency scaling enabled, the difference was \qty{2.6}{\micro\second} and \qty{2.1}{\micro\second} respectively. This indicates that small increases in message size do not  affect the performance of ChaCha20-Poly1305.

The measured encryption time fits well within the limits specified for both Goose and IEC 60834-1, the mean being at most \qty{4.075}{\percent} and the 95th percentile at most \qty{5.6}{\percent} of the time limit for Goose, which has the stronger time limit. With the encryption time of the slowest scenario just under \qty{0.18}{\milli\second} and \qty{0.22}{\milli\second}, respectively, it is similar to what Chelluri et al. found the frame propagation of 12 Goose frames through 10 switches to be, with \SI{1}{\giga\bit\per\second} taking \qty{0.2}{\milli\second} and \SI{100}{\mega\bit\per\second} taking \qty{2}{\milli\second}~\cite{chelluri_integration_2011}.

Nonce generation consistently takes around \qtyrange{2}{12.7}{\micro\second} depending on platform architecture and frequency. The nonce is generated only once per message. Thus, plaintext size does not affect the time of nonce generation.

The binary size is \qty{67}{\kilo\byte} on AArch64 and \qty{19}{\kilo\byte} on x86-64, which is relatively small by the current standard. Further reduction could be possible by tuning WolfSSL, re-implementing ChaCha20 and Poly1305 without WolfSSL, or using the reference implementations from Bernstein~\cite{bernstein_chacha_impl_2008}. This would make it more suitable for embedded or real-time systems.

\section{\label{sec:conclusion_future_works}Conclusion and Future Works}

From these results, we conclude that ChaCha20-Poly1305 is effective for real-time communication and could be used by Goose and Modbus applications, in particular for IoT devices. It is fast, with small messages taking from \qtyrange{52}{224}{\micro\second} to perform IO, encryption, and decryption, showing that it would add around minimally to the communication cycle with encrypting and decrypting \qty{224}{\byte} which includes reading the message twice and all the memory preparation.

This experiment is limited in scope and method. This experiment was performed on a Raspberry Pi and Mini PC  both of which have significantly more computing power and RAM than a PLC such as the s7-1500 with as little as \qty{1.8}{\mega\byte} in memory and a CPU with no advertised clock speed, but which reportedly takes nanoseconds per operation~\cite{siemens_standard_nodate}. However, it also indicates that at least some of the increasing number of  low power IoT devices using Modbus in particular, will be capable of onboard encryption.

Future works could explore modifying Bernstein's reference implementation for development boards with CPU and RAM in the same class as PLCs or on a PLC, as Bernstein's reference implementation is public domain and secure~\cite{bernstein_chacha_impl_2008,bernstein_chacha_2008}. An evaluation of the performance of quantum resistant protocols would also be useful, although this is less of a concern with short term, ephemeral traffic such as ICS control and monitoring.

\begin{credits}
    \subsubsection{\ackname} Support for this work was received from the Government of Iceland's University Collaboration Project (Rannsóknarsetur um netöryggisfræði, Samstarf háskóla) and the European Commissions Digital Europe Programme, Defend Iceland Project (101127307).

    \subsubsection{\discintname}
    The authors have no competing interests to declare that are relevant to the content of this article.

\end{credits}

\bibliographystyle{splncs04}
\bibliography{ref2}

\end{document}

%% file: new_pi.tex
\begin{table}[!ht]
    \centering
    \caption{Mean, 5th and 95th percentiles of execution time for different functions in the ChaCha20-Poly1305 encryption module, with plain text sizes ranging from \qtyrange{28}{224}{\byte} on a Raspberry PI (AArch64) averaged over 100000 runs}
    \begin{tabular}{l l c c c c}
        \toprule
        \multicolumn{2}{l}{Message Size}  & \multicolumn{4}{c}{Time}\\ % \multirow{2}{.7cm}{Message Size} Function
        \cmidrule(lr){3-6}
         & & Functional & Random & Encryption & Decryption\\
        \midrule
        \multirow{3}{.7cm}{\qty{28}{\byte}} & 5th pctl & \qty{134.7}{\micro\second} & \qty{3.4}{\micro\second} & \qty{66.2}{\micro\second} & \qty{63.7}{\micro\second}\\
        &Mean & \qty{154.8}{\micro\second} & \qty{4.3}{\micro\second} & \qty{76.7}{\micro\second} & \qty{73.7}{\micro\second}\\
        &95th pctl & \qty{213.7}{\micro\second} & \qty{5.8}{\micro\second} & \qty{105.9}{\micro\second} & \qty{102.1}{\micro\second}\\
        \midrule

        \multirow{3}{.7cm}{\qty{56}{\byte}} & 5th pctl & \qty{135.2}{\micro\second} & \qty{3.4}{\micro\second} & \qty{66.5}{\micro\second} & \qty{63.9}{\micro\second}\\
        &Mean & \qty{154.5}{\micro\second} & \qty{4.2}{\micro\second} & \qty{76.6}{\micro\second} & \qty{73.5}{\micro\second}\\
        &95th pctl & \qty{214.4}{\micro\second} & \qty{5.8}{\micro\second} & \qty{106.3}{\micro\second} & \qty{102.3}{\micro\second}\\

        \midrule
        \multirow{3}{.7cm}{\qty{112}{\byte}} & 5th pctl & \qty{136.9}{\micro\second} & \qty{3.4}{\micro\second} & \qty{67.1}{\micro\second} & \qty{64.9}{\micro\second}\\
        &Mean & \qty{158.3}{\micro\second} & \qty{4.3}{\micro\second} & \qty{78.2}{\micro\second} & \qty{75.6}{\micro\second}\\
        &95th pctl & \qty{218.4}{\micro\second} & \qty{5.9}{\micro\second} & \qty{107.8}{\micro\second} & \qty{104.9}{\micro\second}\\

        \midrule
        \multirow{3}{.7cm}{\qty{224}{\byte}} & 5th pctl & \qty{139.9}{\micro\second} & \qty{3.4}{\micro\second} & \qty{68.7}{\micro\second} & \qty{66.4}{\micro\second}\\
        &Mean & \qty{162.9}{\micro\second} & \qty{4.3}{\micro\second} & \qty{80.5}{\micro\second} & \qty{77.9}{\micro\second}\\
        &95th pctl & \qty{224.2}{\micro\second} & \qty{5.9}{\micro\second} & \qty{110.4}{\micro\second} & \qty{107.8}{\micro\second}\\

        \bottomrule
    \end{tabular}
    \label{tab:chacha_time_arm}
\end{table}

%% file: new_x64.tex
\begin{table}[!ht]
    \centering
    \caption{Mean, 5th and 95th percentiles of execution time for different functions in the ChaCha20-Poly1305 encryption module, with plain text sizes ranging from \qtyrange{28}{224}{\byte} on a Mini PC (x86-64) averaged over 100000 runs}
    \begin{tabular}{l l c c c c}
        \toprule
        \multicolumn{2}{l}{Message Size}  & \multicolumn{4}{c}{Time}\\
        \cmidrule(lr){3-6}
         & & Functional & Random & Encryption & Decryption\\
        \midrule
        \multirow{3}{.7cm}{\qty{28}{\byte}} & 5th pctl & \qty{51.9}{\micro\second} & \qty{2.0}{\micro\second} & \qty{25.9}{\micro\second} & \qty{22.1}{\micro\second}\\
        & Mean & \qty{57.1}{\micro\second} & \qty{2.4}{\micro\second} & \qty{28.9}{\micro\second} & \qty{25.7}{\micro\second}\\
        & 95th pctl & \qty{62.0}{\micro\second} & \qty{4.5}{\micro\second} & \qty{31.8}{\micro\second} & \qty{28.2}{\micro\second}\\
        \midrule
        \multirow{3}{.7cm}{\qty{56}{\byte}} & 5th pctl & \qty{52.1}{\micro\second} & \qty{2.0}{\micro\second} & \qty{26.0}{\micro\second} & \qty{22.1}{\micro\second}\\
        & Mean & \qty{57.4}{\micro\second} & \qty{2.4}{\micro\second} & \qty{29.1}{\micro\second} & \qty{25.8}{\micro\second}\\
        & 95th pctl & \qty{61.9}{\micro\second} & \qty{4.5}{\micro\second} & \qty{31.8}{\micro\second} & \qty{28.2}{\micro\second}\\
        \midrule
        \multirow{3}{.7cm}{\qty{112}{\byte}} & 5th pctl & \qty{53.0}{\micro\second} & \qty{2.0}{\micro\second} & \qty{26.6}{\micro\second} & \qty{22.4}{\micro\second}\\
        & Mean & \qty{58.5}{\micro\second} & \qty{2.4}{\micro\second} & \qty{29.6}{\micro\second} & \qty{26.3}{\micro\second}\\
        & 95th pctl & \qty{62.8}{\micro\second} & \qty{4.5}{\micro\second} & \qty{32.4}{\micro\second} & \qty{28.6}{\micro\second}\\
        \midrule
        \multirow{3}{.7cm}{\qty{224}{\byte}} & 5th pctl & \qty{54.3}{\micro\second} & \qty{2.0}{\micro\second} & \qty{27.2}{\micro\second} & \qty{23.0}{\micro\second}\\
        & Mean & \qty{59.7}{\micro\second} & \qty{2.4}{\micro\second} & \qty{30.2}{\micro\second} & \qty{26.9}{\micro\second}\\
        & 95th pctl & \qty{64.1}{\micro\second} & \qty{4.5}{\micro\second} & \qty{33.0}{\micro\second} & \qty{29.3}{\micro\second}\\
        \bottomrule
    \end{tabular} 
    \label{tab:chacha_time_x86}
\end{table}

%% file: new_base.tex
\begin{table}[!ht]
    \centering
    \caption{Mean, 5th and 95th percentiles of execution time for different functions in the ChaCha20-Poly1305 encryption module, with plain text sizes ranging from \qtyrange{28}{224}{\byte} on the Mini PC (x86-64) averaged over 100000 runs locked at base clock (\qty{800}{\mega\hertz})}
    \begin{tabular}{l l c c c c}
        \toprule
        \multicolumn{2}{l}{Message Size}  & \multicolumn{4}{c}{Time}\\\cmidrule(lr){3-6}
         & & Functional & Random & Encryption & Decryption\\\midrule
        \multirow{3}{.7cm}{\qty{28}{\byte}} & 5th pctl & \qty{147.3}{\micro\second} & \qty{5.6}{\micro\second} & \qty{73.8}{\micro\second} & \qty{62.0}{\micro\second}\\
        & Mean & \qty{163.3}{\micro\second} & \qty{7.0}{\micro\second} & \qty{83.0}{\micro\second} & \qty{73.1}{\micro\second}\\
        & 95th pctl & \qty{185.0}{\micro\second} & \qty{12.5}{\micro\second} & \qty{90.0}{\micro\second} & \qty{78.8}{\micro\second}\\\midrule

        \multirow{3}{.7cm}{\qty{56}{\byte}} & 5th pctl & \qty{148.5}{\micro\second} & \qty{5.6}{\micro\second} & \qty{74.3}{\micro\second} & \qty{62.7}{\micro\second}\\
        & Mean & \qty{164.9}{\micro\second} & \qty{7.0}{\micro\second} & \qty{83.5}{\micro\second} & \qty{74.1}{\micro\second}\\
        & 95th pctl & \qty{194.7}{\micro\second} & \qty{12.6}{\micro\second} & \qty{90.8}{\micro\second} & \qty{79.6}{\micro\second}\\\midrule

        \multirow{3}{.7cm}{\qty{112}{\byte}} & 5th pctl & \qty{150.6}{\micro\second} & \qty{5.6}{\micro\second} & \qty{75.5}{\micro\second} & \qty{63.6}{\micro\second}\\
        & Mean & \qty{167.3}{\micro\second} & \qty{7.1}{\micro\second} & \qty{84.9}{\micro\second} & \qty{75.2}{\micro\second}\\
        & 95th pctl & \qty{200.4}{\micro\second} & \qty{12.7}{\micro\second} & \qty{92.0}{\micro\second} & \qty{80.6}{\micro\second}\\\midrule

        \multirow{3}{.7cm}{\qty{224}{\byte}} & 5th pctl & \qty{154.9}{\micro\second} & \qty{5.6}{\micro\second} & \qty{77.7}{\micro\second} & \qty{65.7}{\micro\second}\\
        & Mean & \qty{171.8}{\micro\second} & \qty{7.1}{\micro\second} & \qty{87.3}{\micro\second} & \qty{77.2}{\micro\second}\\
        & 95th pctl & \qty{207.0}{\micro\second} & \qty{12.6}{\micro\second} & \qty{94.3}{\micro\second} & \qty{82.8}{\micro\second}\\\bottomrule
    \end{tabular}
    
    \label{tab:chacha_time_x86_lock}
\end{table}